# What lies between a free adiabatic expansion and a quasi-static one?

**E. N. Miranda**[*]

Área de Ciencias Exactas
CCT - CONICET - Mendoza
Sede CRICYT
5500 – Mendoza, Argentina
and
Departamento de Física
Universidad Nacional de San Luis
5700 – San Luis, Argentina

**Abstract:**
An expression is found that relates the initial and final volumes and temperatures for any adiabatic process. It is given in terms of a parameter $r$ that smoothly interpolates between a free adiabatic expansion ($r = 0$) and a quasi-static one ($r = 1$). The parameter has to be evaluated numerically, but an approximate expression is given.

[*] E-mail: emiranda@lab.cricyt.edu.ar

It is a common topic that thermodynamics deals with the macroscopic description of some phenomena while statistical mechanics studies the same phenomena from a microscopic perspective. Ideal gases are a good example of this double approach since many thermodynamical properties can be deduced using the kinetic theory of gases. In a previous publication [1] it has been shown how to obtain the law of an adiabatic expansion or compression from molecular considerations. In this article, we generalize that approach when the expansion is non-reversible. In thermodynamics textbooks [2, 3] two cases are studied: the free expansion of a gas (Joule experiment) that is a highly irreversible process, and the reversible expansion due to a quasi-static process. Now we want to analyze an intermediate situation: the expansion takes place at a high speed, therefore the quasi-static approximation does not hold, but not as high as in a free expansion. In a free expansion the volume changes instantaneously, i.e. with an infinite velocity, while in a quasi-static process the volume change velocity is almost zero. We will find a parameter $r$ that describes the process when the velocity is neither zero nor infinite. The article is written at a level that can be understood by an advanced student with a background in thermodynamics and kinetic theory.

The experimental situation is shown in Figure 1. An ideal monatomic gas is confined in an adiabatic cylinder with cross section $A$. A piston initially at position $x_i$ moves towards the right with a velocity $w$ until it reaches the final position $x_f$. There are two limit situations that can be easily solved. If the piston speed is very high ($w \to \infty$), the ideal gas expands freely. In this case, the initial ($T_i$) and final ($T_f$) temperatures are the same and are unrelated to the volume change. We can write [2, 3]:

$$\frac{T_f}{T_i} = 1 = \left(\frac{V_i}{V_f}\right)^0 \tag{1}$$

On the other hand, if the piston moves extremely slowly ($w \to 0$) the process is quasi-static, and initial temperature and volume ($T_i$, $V_i$) are related to the final ones ($T_f$, $V_f$) by [2, 3]:

$$\frac{T_f}{T_i} = \left(\frac{V_i}{V_f}\right)^{R/C_V} \tag{2}$$

As usual, $R$ is the universal constant of gases and $C_v$ is the molar specific heat at constant volume.

Comparing Eqs. (1) and (2), one is tempted to rewrite them and introduce a parameter $r$ that somehow measures reversibility. In this way, it is proposed that in any adiabatic process holds:

$$\frac{T_f}{T_i} = \left(\frac{V_i}{V_f}\right)^{r\frac{R}{C_V}} \qquad 0 \leq r \leq 1 \tag{3}$$

If the piston moves very slowly, the process is reversible, $r = 1$ and the exponent is $R/C_V$. If the piston velocity is extremely high, the gas expands freely, $r = 0$ and the exponent is $0$. $r$ is a measure of the process reversibility. Our aim in this article is to evaluate $r$ when the piston moves with an intermediate velocity $w$. Thermodynamics cannot be used and we should analyze the problem from a microscopic point of view using the kinetic theory of gases.

Now we focus our attention on Figure 2a. Let us call the molecular density as $\delta = N/V$. The number of collisions with the right wall (i.e. the piston) per unit time $\Delta t$ is:

$$\text{number of collisions}: \quad A.\delta.v_x.\Delta t$$
$$\text{frecuency of collisions} = \text{number of collisions}/\Delta t = A.\delta.v_x$$

The next step is to determine the collisions frequency when the piston moves towards the right with a velocity $w$ as shown in Figure 2b. One should notice that there is no such a thing as a "free compression"; therefore it is senseless to considerer that the piston moves to the left. The number of molecules that hit the piston in the time interval $\Delta t$ is less than the previous situation and it is given by:

$$\text{number of collisions}: \quad A.\delta.(v_x - w).\Delta t$$
$$\text{frecuency of collisions} = \text{number of collisions}/\Delta t = A.\delta.(v_x - w)$$

In each collision the energy and the momentum are conserved magnitudes. If $m$ is the mass of a gas molecule and $M$ that of the piston, one can write:

$$\tfrac{1}{2}m(v_x)^2 + \tfrac{1}{2}Mw^2 = \tfrac{1}{2}m(v_x')^2 + \tfrac{1}{2}M(w')^2 \tag{4}$$

$v_x'$ and $w'$ are respectively the molecular speed in the $x$ direction and the piston speed after the collision.

On the other hand, linear momentum conservation leads us to:

$$mv_x + Mw = mv_x' + Mw' \quad (5)$$

From (4) and (5):

$$v_x' = \left[-v_x\left(1 - \frac{m}{M}\right) + 2w\right]\left(1 + \frac{m}{M}\right)^{-1} \quad (6a)$$

and:

$$w' = \left[2v_x\frac{m}{M} + w\left(1 - \frac{m}{M}\right)\right]\left(1 + \frac{m}{M}\right)^{-1} \quad (6b)$$

Equations (6) can be simplified since $\frac{m}{M} \cong 0$; therefore:

$$v_x' = -v_x + 2w \quad (7a)$$

$$w' = w \quad (7b)$$

Notice that:

$$(v_x')^2 = (v_x)^2 - 4wv_x + 4w^2 \quad (8)$$

If the piston motion were quasi-static, one could asume that $w^2 \cong 0$, however in this paper we are interested in the case that $w \neq 0$, and no term can be negleted in Eq. (8).

The energy change in each collision is:

$$\Delta\varepsilon = \tfrac{1}{2}m(v_x')^2 - \tfrac{1}{2}m(v_x)^2 = 2mw(w - v_x) \quad (9)$$

Taking into account the collisions frequency, the total energy change $\Delta\varepsilon_T$ per unit time due to molecules with a speed $v_x$ is:

$$\frac{\Delta\varepsilon_T}{\Delta t} = A\frac{N}{V}2mw(w - v_x)^2 \quad (10)$$

Since there is a velocity distribution function $f(v_x)$, we should integrate over the whole velocity range to get the total energy change. The kinetic theory of gases [2] states that the probability that a particle has a velocity in the interval $v_x + dv_x$ is:

$$f(v_x) = \left(\frac{m}{2\pi k_B T}\right)^{1/2} \exp\left(-mv_x^2 / 2k_B T\right) \quad (11)$$

As usual $k_B$ is the Boltzman constant. The total energy change $dE$, due to the collisions of particles with any velocity $v_x$, is given by:

$$\frac{dE}{dt} = \int_w^\infty dv_x \ A\frac{N}{V} 2mw(w-v_x)^2 \left(\frac{m}{2\pi k_B T}\right)^{1/2} \exp\left(-mv_x^2/2k_B T\right) \quad (12)$$

Notice that the integral lower limit is $w$ because the particles with $v_x < w$ cannot hit the piston.

It is useful to rewrite Eq. (12) in terms of dimensionless variables $\alpha$ y $\beta$:

$$\begin{aligned} v_x &= \alpha \ v_{rms} \\ w &= \beta \ v_{rms} \\ v_{rms} &= \sqrt{\frac{3k_B T}{m}} \end{aligned} \quad (13)$$

With these new variables, Eq. (12) becomes:

$$\frac{dE}{dt} = \sqrt{\frac{6}{\pi}} \frac{ANm}{V} v_{rms}^3 \beta \int_\beta^\infty (\beta-\alpha)^2 \exp\left(-\frac{3}{2}\alpha^2\right) d\alpha \quad (14)$$

For the sake of simplicity, the integral will be called $I(\beta)$.

The gas volume $V$ can be related to $dt$ as follow:

$$\begin{aligned} V &= A(x_i + \omega t) \\ dV &= A\omega \, dt \\ &= Av_{rms}\beta \, dt \end{aligned} \quad (15)$$

If $n$ is the moles number in the gas, M the molecular weight –do not confuse with the piston mass $M$– , one can write $mN = nM$.

Taking into account all the previous considerations, Eq. (14) becomes:

$$\frac{dE}{dV} = -\sqrt{\frac{6}{\pi}} \frac{nM}{V} v_{rms}^2 I(\beta) \quad (16)$$

At this point a very strong supposition has to be made: the system is clearly out of equilibrium; however the relation $dE = nC_V dT$ between the internal energy E and the molar specific heat $C_V$ is assumed to be valid. What is the rationale behind this assumption? Certainly the whole system is out of equilibrium, however it is assumed that in small spatial regions the gas can be considered to be in equilibrium and the usual thermodynamics relations hold. Of course these regions are small compared with the physical dimensions of the system but large compared with the molecular free path and the atomistic nature of matter is not yet relevant.

Using the expression for $v_{rms}$ given in (13), Eq. (16) can be written as:

$$\frac{dT}{dV} = -\sqrt{\frac{54}{\pi}\frac{R}{C_V}\frac{T}{V}}I(\beta) \tag{17}$$

Our work is almost done. The reversibility coefficient $r$ introduced in Eq. (5) is:

$$r = \sqrt{\frac{54}{\pi}}\int_{\beta}^{\infty}(\beta-\alpha)^2 \exp\left(-\frac{3}{2}\alpha^2\right)d\alpha \tag{18}$$

And we get:

$$\frac{dT}{T} = -r\frac{R}{C_V}\frac{dV}{V} \tag{19}$$

Integrating between the initial and final conditions, we find:

$$\frac{T_f}{T_i} = \left(\frac{V_i}{V_f}\right)^{r\frac{R}{C_V}} \tag{20}$$

Our task is finished. For any speed $w$ of the piston that goes from an initial condition to a final one, the relation between volumes and temperatures is given by (20). The parameter $r$ is evaluated with (18), where $\beta$ is the quotient between the piston velocity and the mean square velocity of the molecules.

Let us evaluate $r$ from (18). If the piston velocity extremely is extremely high, as in a free expansion, then $\beta \to \infty$ and $r = 0$. For a quasi-static process $\beta \to 0$ the integral can be evaluated analytically [4] and $r = 1$.

For intermediate values of $\beta$ one has to calculate $r$ numerically. The results are shown in Figure 3. For $\beta = 1$, i.e. the piston velocity equals to the mean square velocity of the molecules, $r \approx 0$ as in a free expansion. This result is easy to understand: the piston moves as quickly as the molecules and they hardly can hit it.

To have an approximate value of $r$ without evaluating the integral, an exponential has been fitted to the data. The following values have been found:

$$\begin{aligned} r &= \exp(-\beta/\tau) \\ \tau &= 0.31 \end{aligned} \tag{21}$$

Our work is finished. Our aim was to interpolate between two different adiabatic processes. Free expansion and quasi-static expansion can be both analyzed with thermodynamics; the kinetic theory of gases has been used to study the adiabatic processes that lie between those two. Our main result is to write a relation for initial and final volumes and temperatures in any adiabatic process as shown in Eq. (3) and (20). Through a microscopic analysis, an explicit expression has been found for the *r* parameter –see Eq. (18)- that somehow measure how close a process is to a quasi-static one. This parameter has been evaluated numerically -see Figure 3- but its approximate value can be obtained with Eq. (21).

What can we learn from this exercise ? We have proved that the microscopic description of a gas agrees with the phenomenological one given by thermodynamics. The macroscopic approach allows us to evaluate the gas behavior in two extreme situations, while the microscopic one is much more general (but also more cumbersome). The description given by the kinetic theory of gases interpolates smoothly between the two extreme cases that can be solved by thermodynamics. The parameter *r* changes continuously from *0* to *1* and indicates the change from a highly irreversible process –free adiabatic expansion- to a perfectly reversible one –quasistatic adiabatic expansion-.

In addition, it is given a quantitative estimation of the piston velocity needed for a "free" expansion. From Figure 3 it is clear that the piston speed should be slightly greater than the average molecular velocity to have a free-expansion behavior (i.e. *r = 0*). And in physical terms this result is easy to understand: if the piston speed is comparable to those of the molecules, they can hardly hit the piston that moves outwards.

Finally, it has been shown that an equilibrium thermodynamics definition – $dE = C_v dT$ – is also valid in a non-equilibrium situation if the system can be divided in small regions that are locally in equilibrium. This assumption –local thermodynamical equilibrium – is a cornerstone of linear non-equilibrium thermodynamics and has far reaching consequences.

From a pedagogical point of view, this exercise can be understood for a senior physics student and it goes beyond the usual textbooks examples.

**Acknowledment:** The author thanks the National Scientific Research Council of Argentina (CONICET) for financial support.

## Figure captions

### Figure 1:

The experimental situation considered in this article is shown. The cylinder contains a monatomic perfect gas. All the walls are adiabatic, i.e. there is no heat loss or gain. A piston with cross section $A$ moves towards the right with a velocity $w$, starting at $x_i$ and finishing at $x_f$. The initial temperature is $T_i$ and the final one is $T_f$. Obviously, the initial volume is $V_i = A \cdot x_i$, and the final one is $V_f = A \cdot x_f$.

### Figure 2:

Consider the two situations shown in this figure. The aim is to evaluate how many molecules with velocity $v_x$ hit the piston in the time interval $\Delta t$. If the piston is at rest –Fig. 2a- two molecules will hit the piston in that time interval while the third one won't because its initial position is larger than $v_x \Delta t$. In Fig. 2b the piston is moving towards the right with velocity $w$. In this situation only one molecule will collide with the piston. The other two won't because their initial positions are larger that $(v_x - w)\Delta t$. The detailed calculation is done in the main text.

### Figure 3:

The value of the parameter $r$, that somehow measures reversibility, is shown as a function of $\beta$ (the quotient between the piston speed and the mean square root velocity of the gas molecules). For $\beta \geq 1$ the parameter is almost zero, meaning the process has is highly irreversible as in a free expansion.

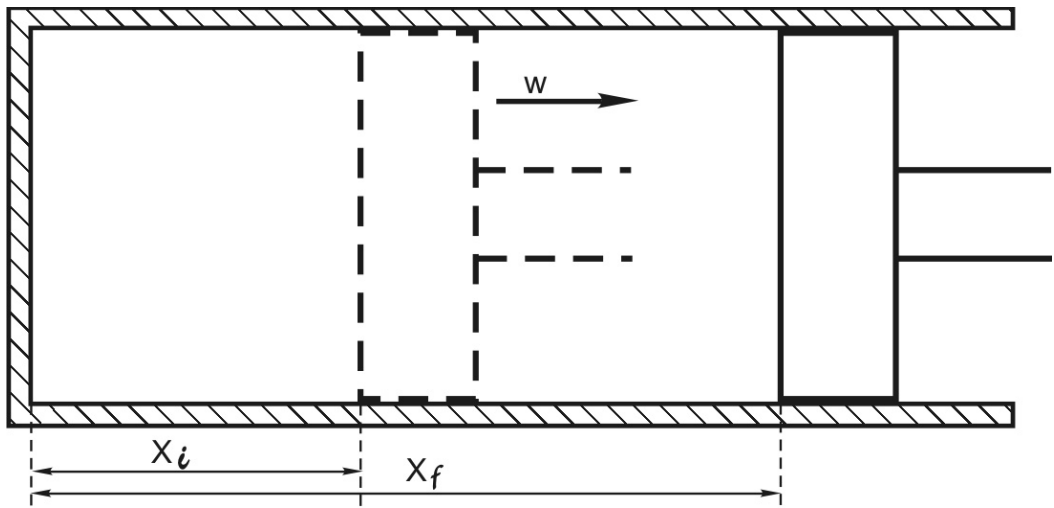

Figure 1

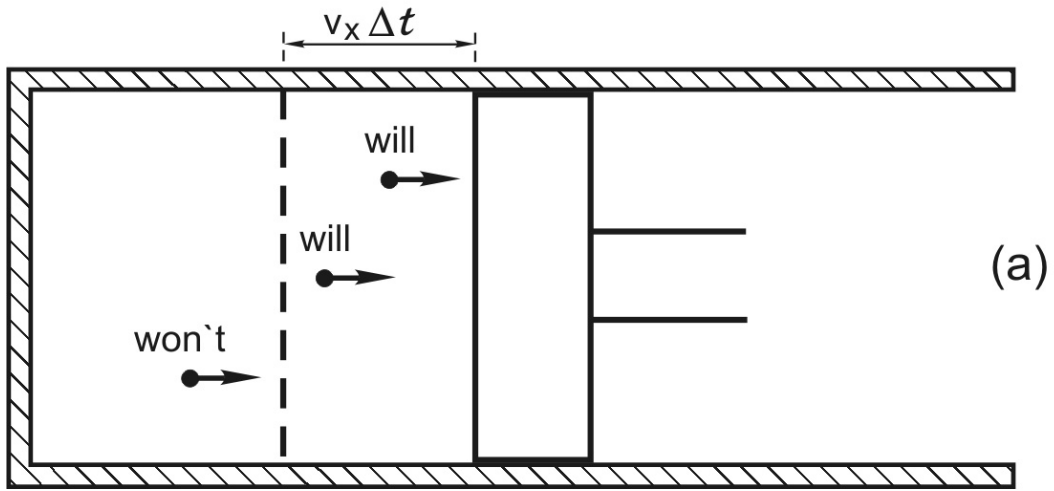
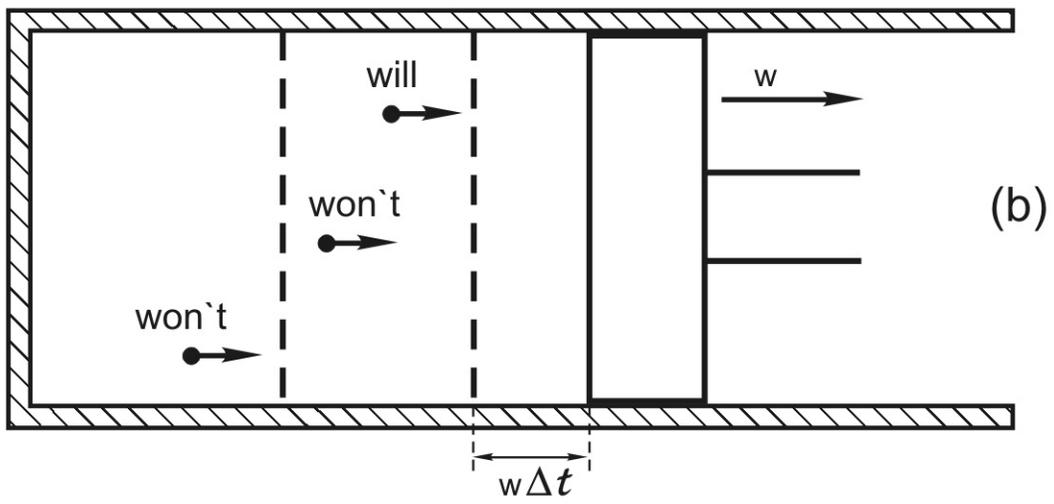

Figure 2

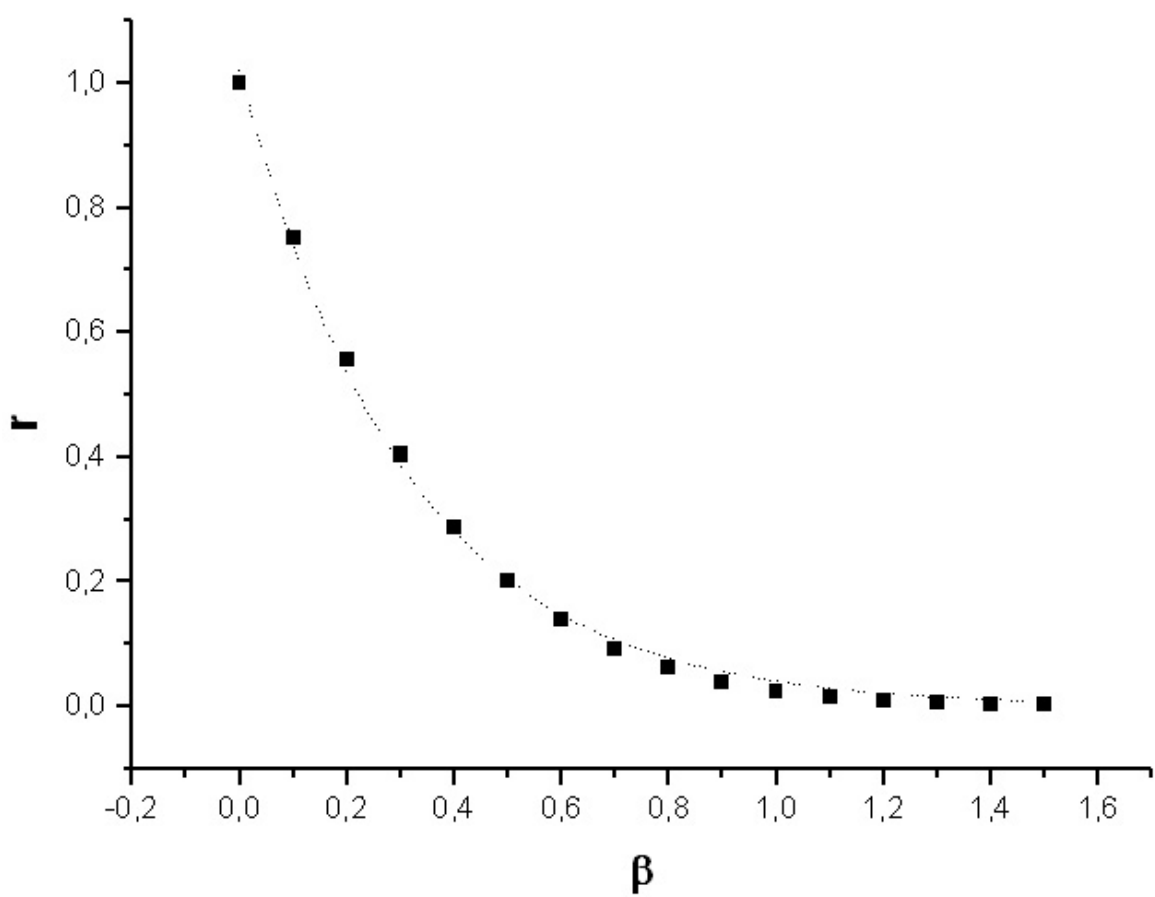

**Figure 3**